\documentclass[12pt]{iopart}

\usepackage{iopams}  
\usepackage{textcomp, graphicx, braket}
\usepackage{booktabs, multirow}
\usepackage{hyperref}
\hypersetup{colorlinks=true, linkcolor=red, urlcolor=magenta}
\newcommand{\abs}[1]{\left| #1 \right|}
\newcommand{\bte}{$\beta$-ensemble}
\newcommand{\brpe}{$\beta$-RPE}
\newcommand{\cbr}[1]{\ensuremath{\left\{ #1 \right\}}}
\newcommand{\del}[1]{\ensuremath{\left( #1 \right)}}
\newcommand{\fbsli}[2]{\ensuremath{\mathrm{I}_{#1}\del{#2}}}

\newcommand{\fdsnf}[1]{\ensuremath{\mathrm{D}_{\mathrm{F}}\del{#1}}}	
\newcommand{\felpe}[1]{\ensuremath{\mathrm{EllipticE}\del{#1}}}

\newcommand{\ferf}[1]{\ensuremath{\mathrm{Erf}\del{#1}}}
\newcommand{\ferfi}[1]{\ensuremath{\mathrm{Erfi}\del{#1}}}
\newcommand{\fgamma}[1]{\ensuremath{\Gamma\del{#1}}}					
\newcommand{\ipr}{\ensuremath{\mathrm{I}}}
\newcommand{\mean}[1]{\ensuremath{\left\langle#1\right\rangle}}
\def\ot{\ensuremath\mathbb{O}}	
\newcommand{\prob}[1]{\ensuremath{\mathrm{P}\del{#1}}}

\newcommand{\ppsn}[1]{\ensuremath{\mathrm{P}_{\mathrm{Poisson}}\del{#1}}}

\newcommand{\pnns}[2]{\ensuremath{\mathrm{P}_{\gamma}^{(#1)}\del{#2}}}
\newcommand{\rpe}{Rosenzweig-Porter Ensemble}
\newcommand{\sbr}[1]{\ensuremath{\left[ #1 \right]}}
\newcommand{\tTh}{\ensuremath{t_{\mathrm{Th}}}}		
\newcommand{\tR}{\ensuremath{t_{\mathrm{R}}}}		

\begin{document}

\title[Dynamical signatures of crossover in $2\times 2$ ensembles]{Dynamical Signatures of Chaos to Integrability Crossover in $2\times 2$ Generalized Random Matrix Ensembles}

\author{Adway Kumar Das \& Anandamohan Ghosh}

\address{Department of Physical Sciences, Indian Institute of Science Education and Research Kolkata, Mohanpur, 741246 India}
\ead{\mailto{akd19rs062@iiserkol.ac.in}, \mailto{anandamohan@iiserkol.ac.in}}
\vspace{10pt}
\begin{indented}
\item[]\today
\end{indented}
\begin{abstract}
We introduce a two-parameter ensemble of generalized $2\times 2$ real symmetric random matrices called the $\beta$-Rosenzweig-Porter ensemble (\brpe), parameterized by $\beta$, a fictitious inverse temperature of the analogous Coulomb gas model, and $\gamma$, controlling the relative strength of disorder. \brpe\ encompasses RPE from all of the Dyson's threefold symmetry classes: orthogonal, unitary and symplectic for $\beta=1,2,4$. 
Firstly, we study the energy correlations by calculating the density and 2nd moment of the Nearest Neighbor Spacing (NNS) and robustly quantify the crossover among 
various degrees of level repulsions. 
Secondly, the dynamical properties are determined from an exact calculation of the temporal evolution of the fidelity enabling an identification of the characteristic Thouless and the equilibration timescales.   
The relative depth of the correlation hole in the average fidelity serves as a dynamical signature of the crossover from chaos to integrability and enables us to construct the phase diagram of \brpe\ in the $\gamma$-$\beta$ plane. 
{Our results are in qualitative agreement with numerically computed fidelity for $N\gg2$ matrix ensembles. Furthermore, we observe that for large $N$ the 2nd moment of NNS and the relative depth of the correlation hole exhibit a second order phase transition at $\gamma=2$.}
\end{abstract}

\vspace{2pc}

\submitto{\JPA}

\section{Introduction}
Random matrix theory provides important insights into the dynamical aspects of a system by analyzing its spectral statistics. For example, an integrable system behaves as an insulator while producing uncorrelated eigenvalues and localized eigenstates as in the Poisson ensemble \cite{Berry3}. Contrarily, Wigner-Dyson ensembles (WDE) \cite{Dyson1, Altland2} can replicate the energy spectra in completely chaotic systems \cite{Bohigas1, Berry8} while respective eigenstates are ergodic \cite{Berry5} leading to a diffusive transport of any excitation in the system.
In a closed quantum system, the energy correlations and the eigenstate localization properties govern the dynamical response of the system, e.g.~the {\it fidelity} of an initially localized state \cite{Gorin2, Torres1}. The time required for the system to equilibrate and the equilibrium value of the fidelity are indicative of the degree of chaos in the system \cite{Torres6}. 
Such dynamical signatures can be measured experimentally and can identify various quantum mechanical phases \cite{Schreiber1, Sutradhar1, Haldar1}. While the integrable and chaotic limits are well characterized, the evidence of intermediate statistics have generated considerable interest in the context of many-body localization transition \cite{Ni1, Rao3, Prosen1, Sutradhar2, Roy2}. This necessiates exploring dynamical signatures of random matrix ensembles beyond the WDE. 

The WDE have two constraints: statistical independence of the matrix elements and canonical invariance \cite{Porter1}. We need to relax either or both of these constraints to capture the intermediate spectral properties found in the systems deviating from the conventional Boltzmann statistics \cite{Pino3, Pino2, Luca1, Garcia2, Ray1, Wang3, Ruidas1}. A prominent random matrix model without canonical invariance is the \rpe\ (RPE). It was originally introduced to explain the atomic spectra of elements like Y, Zr, Nb, Pd \cite{Rosenzweig1}. Since RPE is essentially Poisson ensemble perturbed by WDE, the perturbation strength can be considered as a fictitious time and RPE can be cast as a Brownian ensemble \cite{Pandey2, Pandey8, Shukla1, Mergny1, Biroli4}. Again RPE can be considered as a deformed ensemble \cite{Carvalho1, Carvalho2, Das3, Das4} where the symmetries present in an integrable system are broken. Apart from the obvious appeal as an interpolating random matrix ensemble, interestingly, RPE hosts three distinct phases: ergodic, non-ergodic extended phase having fractal eigenstates and a localized phase \cite{Kravtsov1, Pino1, Soosten1, Monthus1}. Corresponding phase transitions have been explored experimentally in microwave resonators \cite{Zhang3}. The localized and critical states at the Anderson transition point of RPE have the same statistical properties as those in the hierarchical lattices such as Bethe lattice or Random Regular Graph \cite{Kravtsov1}. The Fock space of generic isolated quantum many-body systems has a hierarchical structure \cite{Altshuler4}, hence Anderson localization in RPE can serve as a toy model for the many-body localization. Therefore, RPE has gained a lot of attention in recent times and has been extensively studied \cite{Kravtsov1, Khaymovich1, Khaymovich2, Tomasi1, Bogomolny1, Amini1, Facoetti1, Venturelli1}.

In this work, we focus on the dynamical properties of the $2\times 2$ RPE to identify the crossover from chaos to integrability. The simplicity of $2\times 2$ matrices makes them amenable to exact analytical treatments while capturing the essence of large  system sizes, $N\gg 2$ \cite{Lenz1, Pichard1}. For example, the Wigner's surmise obtained for $2\times 2$ WDE are reasonable approximations of the density of level spacing for large $N$. Apart from their simplicity, 2-level systems are often encountered in various disciplines e.g.~optics \cite{Larson1, Gupta1, Mollow1, Makin1}, wave localization \cite{Berry7}, PT symmetry \cite{Ganainy1, Bender1}, exciton dynamics \cite{Caycedo1}, quantum information processing \cite{Zagoskin1}, beam shifts in quantum optics \cite{Toppel1, Gotte1, Modak2}. The matrix elements of $2\times 2$ RPE follow Gaussian distributions with mean 0 and the variances \cite{Kravtsov1}
\begin{equation}
	\label{eq_H_element}
	\mean{H_{n, n}^2} = 1,\quad \mean{\abs{H_{n, m}}^2} = \frac{\beta}{2^{\gamma+1}},\quad \gamma\in\mathbb{R},
\end{equation}
where $\gamma$ parameterizes the disorder strength and $\beta$ is the Dyson's index characterizing the symmetry class, e.g.~$\beta = 1,2,4$ correspond to real, complex, quaternion matrices from the orthogonal, unitary, symplectic symmetry classes, respectively. For $2\times 2$ matrices from all the three symmetry classes, we compute the spectral statistics and the fidelity of an initially localized state. The ensemble averaged fidelity after an initial quadratic decay jumps and relaxes to an equilibrium value, decreasing with the loss of integrability. The dip, occurring before and below the equilibrium value, is known as the {\it correlation hole}. The relative depth of the correlation hole is denoted by $\kappa$ and is used to quantify the crossover from chaos to integrability which is similar to that inferred from the spectral statistics.  We introduce a two-parameter ($\gamma, \beta$) model called the $\beta$-Rosenzweig-Porter ensemble (\brpe) which  reduces to \bte\ \cite{Dumitriu1} for $\gamma = 0$. 
We derive an analytical expression of the fidelity as a function of time for \brpe\ unifying all the abovementioned symmetry classes.
We numerically obtain a phase diagram of \brpe\ using $\kappa$ to identify the crossover from chaos to integrability in the $\gamma$-$\beta$ plane. Based on the analytical expression of fidelity of $2\times 2$ \bte, we propose an ansatz for the average fidelity of any 2-level systems, therefore providing a dynamical counterpart of the empirical distributions of spectral statistics.
{We have numerically computed fidelity for  $N\gg2$ matrices and show that the relative depth of the correlation hole exhibit a second order phase transition at $\gamma = 2$ for RPE from all the three symmetry classes. In spite of the simplicity, the dynamical signatures of parameter dependent crossover in 2-level system possess qualitative similarity to those of random matrices with $N\gg 2$.}

\section{Orthogonal \rpe ~(O-RPE)}
For a choice of orthogonal eigenvectors, $\ket{\Phi_1} = \left(\begin{array}{c}
	\cos\theta\\
	-\sin\theta
\end{array}\right)$ and $\ket{\Phi_2} = \left(\begin{array}{c}
	\sin\theta\\
	\cos\theta
\end{array}\right)$ with the eigenvalues $E_1$ and $E_2$, respectively,
any $2\times 2$ real symmetric matrix can be expressed as
\begin{equation}
	\label{eq_H_O-RPE}
	H = \left(\begin{array}{cc}
		E_1\cos^2\theta + E_2\sin^2\theta & \frac{1}{2}(E_1-E_2)\sin 2\theta\\
		\frac{1}{2}(E_1-E_2)\sin 2\theta & E_1\sin^2\theta + E_2\cos^2\theta
	\end{array}\right).
\end{equation}
The Jacobian of the transformation from matrix space to eigenspace is the Vandermonde determinant, $E_1-E_2$. Then following equation~\ref{eq_H_element}, we obtain the joint density of eigenvalues and $\theta$ for $2\times 2$ O-RPE as
\begin{equation}
	\label{eq_P_E_theta_ORPE}
	\prob{E_1, E_2, \theta} = \frac{2^{\frac{\gamma}{2}}}{4\pi^{\frac{3}{2}}} |E_1-E_2|e^{ -\frac{E_1^2+E_2^2}{2} + \del{1 - 2^\gamma}\sin^2 2\theta\frac{(E_1-E_2)^2}{4} }.
\end{equation}
The energy correlation can be captured by the Nearest Neighbor Spacing (NNS), $S = \abs{E_1-E_2}$ \cite{Mehta1} and the corresponding density is given by
\begin{equation}
\label{eq_S_def}
\prob{S} = \int_0^{\pi}d\theta \prob{S, \theta}  = \int_0^{\pi} d\theta \int_{-\infty}^{\infty} dE_1dE_2\; \prob{E_1, E_2, \theta}\delta\del{S - |E_1-E_2|}.
\end{equation}
The level spacing is scaled as $s = \frac{S}{\mean{S}}$ to fix the unit of energy (\ref{sec_A_unfold}). Then density of NNS for $2\times 2$ O-RPE is \cite{Kota1, Huu-Tai1, Berry1}
\begin{equation}
	\label{eq_s_ORPE}
	\pnns{1}{s} = 2^{2+\frac{|\gamma|}{2}}f_\gamma se^{-\del{1+2^{|\gamma|}} f_\gamma s^2} \fbsli{0}{\del{1-2^{|\gamma|}} f_\gamma s^2}
\end{equation}
where $f_\gamma = \frac{ \del{\felpe{1-2^{-|\gamma|}}}^2 }{2\pi}$ and $\felpe{x} = \int_{0}^{\frac{\pi}{2}} d\theta\:\sqrt{1 - x\sin^2\theta}$ is the complete elliptic integral. In figure~\ref{fig_ORPE}(a), $\pnns{1}{s}$ is shown for various $\gamma$ where $\gamma = 0$ corresponds to the Gaussian Orthogonal Ensemble (GOE). The Taylor expansion of $\pnns{1}{s}$ around $s = 0$ apparently implies linear level repulsion irrespective of $\gamma$. However, $\pnns{1}{s}$ attains the maximum around $s^\ast(\gamma) = \del{3f_\gamma\del{1+2^{|\gamma|}}}^{-\frac{1}{2}}$ where $s^\ast(\gamma) \to 0$ in the integrable limit $\gamma\gg 1$ and  $\pnns{1}{s} \to \ppsn{s} = \frac{2}{\pi}e^{-\frac{s^2}{\pi}}$. We can fit $\pnns{1}{s}$ with empirical distributions (e.g.~Brody \cite{Brody1}, Berry-Robnik \cite{Berry2}, Izrailev \cite{Izrailev1}) to estimate the degree of level repulsion. Nevertheless, such a numerical fit captures the global shape of $\prob{s}$ without necessarily reflecting the behavior of $\prob{s\ll 1}$ \cite{Sorathia1}. To avoid such ambiguity, we compute $\mean{s^2}$, the 2nd moment of NNS, to quantify the exact crossover from level clustering to repulsion. The functional form of $\mean{s^2}$ is given in table~\ref{tab_s2} and plotted in the inset of figure~\ref{fig_ORPE}(a). Note that $\mean{s^2}$ is minimum at and symmetric about $\gamma = 0$ (i.e.~the GOE limit).

\newcommand{\x}{0.85}
\begin{figure}[t]
	\raggedleft
	\includegraphics[width=\x\textwidth]{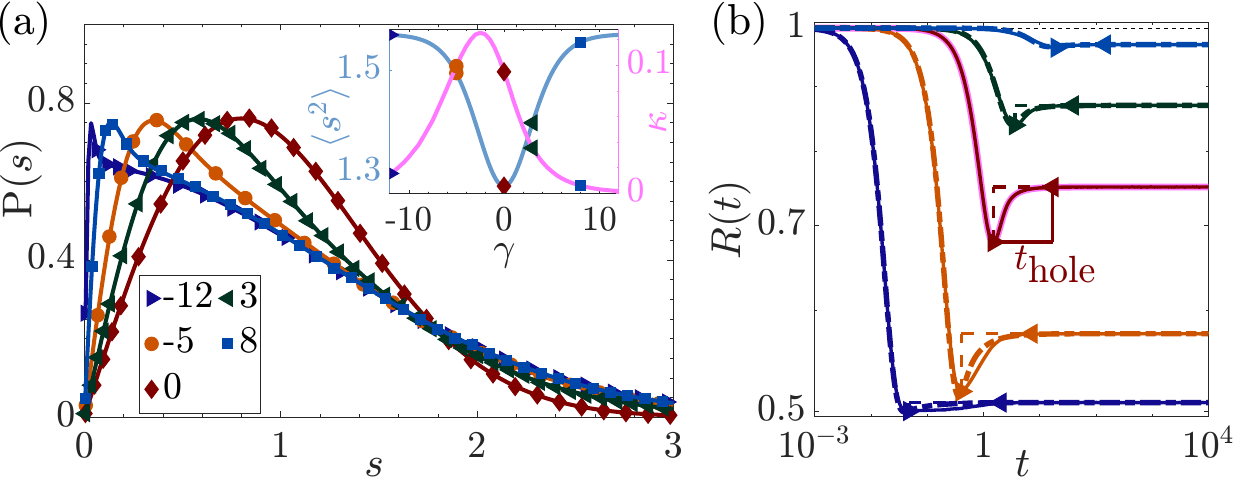}
	\caption{O-RPE: (a) density of NNS for various $\gamma$. Solid lines denote analytical form in equation~\ref{eq_s_ORPE} wheres the markers denote simulation over $10^7$ samples. Inset shows $\mean{s^2}$, 2nd moment of NNS, where the blue solid line denotes analytical form in table~\ref{tab_s2} and $\kappa$, the relative depth of the correlation hole in the fidelity (pink solid line).
		(b) temporal evolution of fidelity where Thouless and relaxation times are denoted by markers and correlation hole is marked by dotted lines. Pink bold line is the analytical form for GOE (table~\ref{tab_WD_S_Rt}) and the dashed lines are fit using equation~\ref{eq_Rt_fit}. The asymptotic values are given in table~\ref{tab_R_bar}.
	}
	\label{fig_ORPE}
\end{figure}
Now we investigate the dynamical properties of $2\times 2$ O-RPE. Given the eigenvectors $\ket{\Phi_1}$ and $\ket{\Phi_2}$ with non-degenerate energies $E_1$ and $E_2$, respectively, we initialize a 2-level system governed by $H$ in any arbitrary initial state $\ket{\Psi_0}\neq \ket{\Phi_{1, 2}}$. Then fidelity of the initial state has a periodic time evolution 
\begin{equation}
	\label{eq_Rt_general_2}
	R(t)\equiv \abs{ \braket{\Psi_0|e^{-iHt}|\Psi_0} }^2 = 1 - 4c_1^2c_2^2\sin^2\frac{S t}{2}
\end{equation}
where $c_j = \abs{ \braket{\Phi_j|\Psi_0} }$ i.e.~modulus of the overlap of $j$th eigenstate with $\ket{\Psi_0}$ and $S = \abs{E_2-E_1}$ determines the periodicity of Rabi oscillation \cite{Merlin1}. The fidelity of a general initial state, $\ket{\Psi_0} = \left(\begin{array}{c}
	\cos\psi\\ \sin\psi
\end{array}\right)$, is studied in \ref{sec_A_gen_Rt}. Since we want to identify the crossover from integrability to chaos, let us choose $\ket{\Psi_0}$ as one of the eigenstates in the integrable limit, i.e.~$\ket{\Psi_0} = \left(\begin{array}{c}
	1\\ 0
\end{array}\right)$. Then for general $2\times 2$ real symmetric matrices, we can express the fidelity as
\begin{equation}
	\label{eq_Rt_general}
	  R(t) = 1 - \frac{\sin^2 2\theta}{2}\del{1 - \cos S t}	
\end{equation}
and its ensemble average is given by
\begin{equation}
	\label{eq_Rt_general8}
	\mean{R(t)} = \overline{R} + \int dS d\theta \frac{\sin^2 2\theta}{2} \prob{S, \theta} \cos \del{S t}.
\end{equation}
The second term in $\mean{R(t)}$ vanishes for $t\gg 1$, hence ensemble averaging suppresses the Rabi oscillation and the fidelity equilibrates to
\begin{equation}
\overline{R} = 1 - \frac{1}{2}\int dS d\theta \prob{S, \theta} \sin^2 2\theta.
\end{equation}
It is to be noted that $\overline{R}$ is equivalent to the ensemble averaged Inverse Participation Ratio (IPR, $\ipr = \sum_j |\Phi(j)|^4$).
\begin{table}[t]
	\raggedleft
	\begin{tabular}{|c|ccccc|}
		\hline
		Model & \multicolumn{1}{c|}{$\mean{s^2}$}	& \multicolumn{1}{c|}{$\gamma\to -\infty$} 	& \multicolumn{1}{c|}{$\gamma\to 0$}	& \multicolumn{1}{c|}{$\gamma\to \infty$}	& $\gamma\to 0, N\to \infty$ \\ \hline
		O-RPE & \multicolumn{1}{c|}{$\frac{\pi\del{1 + 2^{-|\gamma|}}}{2 \del{\felpe{1-2^{-|\gamma|}}}^2}$}			& \multicolumn{1}{c|}{$\frac{\pi}{2}$}    & \multicolumn{1}{c|}{$\frac{4}{\pi}\approx 1.273$}     & \multicolumn{1}{c|}{$\frac{\pi}{2}$}   & 1.285 \\ \hline
		U-RPE & \multicolumn{1}{c|}{$\frac{\pi\del{1 + 2^{\gamma-1}}}{\del{2^{\frac{\gamma}{2}} + c_\gamma}^2}$} 	& \multicolumn{1}{c|}{$\frac{4}{\pi}$}    & \multicolumn{1}{c|}{$\frac{3\pi}{8}\approx 1.178$}    & \multicolumn{1}{c|}{$\frac{\pi}{2}$}   & 1.180 \\ \hline
		S-RPE & \multicolumn{1}{c|}{$\frac{|1-2^\gamma|(4+2^\gamma)}{2^{\gamma+1}\mu^2}$}  & \multicolumn{1}{c|}{$\frac{3\pi}{8}$}   & \multicolumn{1}{c|}{$\frac{45\pi}{128}\approx 1.104$} & \multicolumn{1}{c|}{$\frac{\pi}{2}$}   & 1.104 \\ \hline
		\multicolumn{1}{|c|}{\brpe} & \multicolumn{5}{c|}{$\mean{s^2} = \frac{2 + 4\beta\sigma^2}{\mu^2},\quad \mu = \frac{ \beta \fgamma{\frac{\beta}{2}} }{\sigma^\beta 2^{\frac{\beta}{2}} \fgamma{\frac{\beta+1}{2}} } {}_2F_1\del{\frac{\beta}{2}, \frac{\beta+2}{2}; \frac{\beta+1}{2}; 1-\frac{1}{2\sigma^2}}$} \\ \hline
		\multicolumn{1}{|c|}{\bte} & \multicolumn{5}{c|}{$\mean{s^2} = \frac{\del{\beta+1}\fgamma{\frac{\beta+1}{2}}^2 }{2\fgamma{1+\frac{\beta}{2}}^2}$} \\ \hline
	\end{tabular}
	\caption{2nd moment of NNS for $2\times 2$ RPE from various symmetry classes. For S-RPE, $\mu$ is given in equation~\ref{eq_s_SRPE}. The last column shows $\mean{s^2}$ for the WDE in the thermodynamic limit ($N\to\infty$), obtained by interpolating numerically evaluated $\mean{s^2}$ from the middle 50\% spectrum up to system size, $N = 32768$. For Poisson ensemble, $\mean{s^2}_{N = 2} = \frac{\pi}{2}$ and $\mean{s^2}_{N\to\infty} = 2$.}
	\label{tab_s2}
\end{table}
For $2\times 2$ O-RPE, we use the distribution from equation~\ref{eq_P_E_theta_ORPE} in equation~\ref{eq_Rt_general8} to get the ensemble averaged fidelity 
\begin{equation}
	\label{eq_Rt_ORPE}
	\eqalign{ &\mean{R(t)} = \overline{R} + 2^{\frac{\gamma}{2}-3}\int_{0}^{\infty} dS\; \cos\del{St} f(S)\\
			  &f(S) = S e^{ \del{c_\gamma - \frac{1}{4}} S^2 } \sbr{ \fbsli{0}{c_\gamma S^2} + \fbsli{1}{c_\gamma S^2} },\quad -\infty< c_\gamma = \frac{1-2^\gamma}{8}<\frac{1}{8}.
			}
\end{equation}
$\mean{R(t)}$ is evaluated for different values of $\gamma$ and is shown in figure~\ref{fig_ORPE}(b) and the equilibrium value $\overline{R}$ is listed in table~\ref{tab_R_bar}. Equation~\ref{eq_Rt_ORPE} implies that for very short time, fidelity exhibits a quadratic decay, $\mean{R(t)} \sim 1 - \frac{t^2}{2^{\gamma+1}}$. After attaining the minimum at Thouless time, $\tTh$, the fidelity grows and asymptotically reach the equilibrium value $\overline{R}$ at the relaxation time, $\tR$. The gap between the Thouless and the relaxation time is known as the correlation hole. The relative depth of the correlation hole, $\kappa \equiv 1 - \frac{ \mean{R(\tTh)} }{\overline{R}}$, reflects the long-range energy correlations: $\kappa = 0$ and $\kappa > 0$ implies uncorrelated and correlated spectrum, respectively \cite{Torres1}. In the inset of figure~\ref{fig_ORPE}(a), we plot $\kappa$ as a function of $\gamma$ showing a crossover from integrable to chaotic regime. Note that $\kappa$ does not show maximum at $\gamma = 0$ but qualitatively reciprocates the crossover observed in $\mean{s^2}$ derived from the eigenvalues only. We also observe that both $\tTh$ and $\tR$ increase exponentially as a function of $\gamma$, i.e.~$\mean{R(t)}$ takes longer time to equilibrate while the equilibrium value itself approaches unity as we go towards the integrable limit. Consequently monitoring the fidelity, we can identify the crossover from chaos to integrability in case of $2\times 2$ O-RPE. Such a crossover can be equivalently understood as a deformation of the symmetry present in an integrable Hamiltonian (\ref{sec_A_deformed}). The fidelities in case of GOE for $N = 2$ and $N\gg 1$ are compared in \ref{sec_A_Rt_compare}. 
\begin{table}[t]
	\raggedleft
	\begin{tabular}{|c|cccc|}
			\hline
			Model                       & \multicolumn{1}{c|}{$\overline{R}$} & \multicolumn{1}{c|}{$\gamma\to-\infty$} & \multicolumn{1}{c|}{$\gamma\to0$}  & $\gamma\to\infty$ \\ \hline
			O-RPE                       & \multicolumn{1}{c|}{$1 - \frac{1}{2\del{1+2^{\frac{\gamma}{2}}}}$}             & \multicolumn{1}{c|}{$\frac{1}{2}$}      & \multicolumn{1}{c|}{$\frac{3}{4}$} & 1                  \\ \hline
			U-RPE                       & \multicolumn{1}{c|}{$\frac{\pi\del{1 + 2^{\gamma-1}}}{\del{2^{\frac{\gamma}{2}} + c_\gamma}^2}$}             & \multicolumn{1}{c|}{$\frac{1}{2}$}      & \multicolumn{1}{c|}{$\frac{2}{3}$} & 1                  \\ \hline
			S-RPE                       & \multicolumn{1}{c|}{$\frac{4^{\gamma+1} + 2 - 3 \del{2^\gamma + 4^\gamma \frac{\mathrm{sech}^{-1}2^\frac{\gamma}{2} }{\sqrt{1 - 2^\gamma}}} }{4\del{2^\gamma - 1}^2}$}             & \multicolumn{1}{c|}{$\frac{1}{2}$}      & \multicolumn{1}{c|}{$\frac{3}{5}$} & 1                  \\ \hline
			\multicolumn{1}{|c|}{\brpe} & \multicolumn{4}{c|}{$1 - {}_2F_1\del{\frac{\beta+1}{2}, \frac{\beta+2}{2}; \frac{\beta+3}{2}; 1-2^\gamma } \frac{\beta2^{\frac{\beta\gamma}{2} - 1}}{ \beta+1 }$} \\ \hline
			\multicolumn{1}{|c|}{\bte} & \multicolumn{4}{c|}{$1 - \frac{\beta}{2\del{\beta+1}}$} \\ \hline
		\end{tabular}
	\caption{Asymptotic value of fidelity for $2\times 2$ RPE from various symmetry classes along with limiting cases of $\gamma$. For U-RPE, $c_\gamma$ is given in table~\ref{tab_s_URPE}.
	}
	\label{tab_R_bar}
\end{table}
\section{Unitary \rpe ~(U-RPE)}
Any $2\times 2$ complex hermitian matrix can be expressed as
\begin{equation}
	\label{eq_H_U-RPE}
	H = \left( \begin{array}{cc}
		E_1\cos^2\theta + E_2\sin^2\theta& e^{i\phi}\frac{E_1-E_2}{2}\sin 2\theta\\
		e^{-i\phi}\frac{E_1-E_2}{2}\sin 2\theta& E_2\cos^2\theta + E_1\sin^2\theta
	\end{array} \right)
\end{equation}
where the eigenvectors are $\ket{\Phi_1} = \left( \begin{array}{c}
	\cos\theta e^{-i\phi}\\
	-\sin\theta
\end{array} \right)$ and $\ket{\Phi_2} = \left( \begin{array}{c}
	\sin\theta\\
	\cos\theta e^{i\phi}
\end{array} \right)$ ignoring a global phase factor of the form $e^{i\alpha}$. The Jacobian of transformation from matrix space to eigenspace is $\frac{\sin 2\theta}{2}(E_1-E_2)^2$. Proceeding as in the previous section, i.e.~starting with equation~\ref{eq_H_element} and integrating over $\phi\in\sbr{0, \frac{\pi}{2}}$, we obtain the joint density of eigenvalues and $\theta$ for $2\times 2$ U-RPE
\begin{equation}
	\label{eq_P_E_theta_URPE}
	\prob{E_1, E_2, \theta} = \frac{2^\gamma}{8\pi}(E_1-E_2)^2 \abs{\sin 2\theta}e^{ -\frac{E_1^2+E_2^2}{2} + \del{1 - 2^\gamma}\frac{(E_1-E_2)^2}{4}\sin^2 2\theta}.
\end{equation}
\begin{figure}[t]
	\raggedleft
	\includegraphics[width=\x\textwidth]{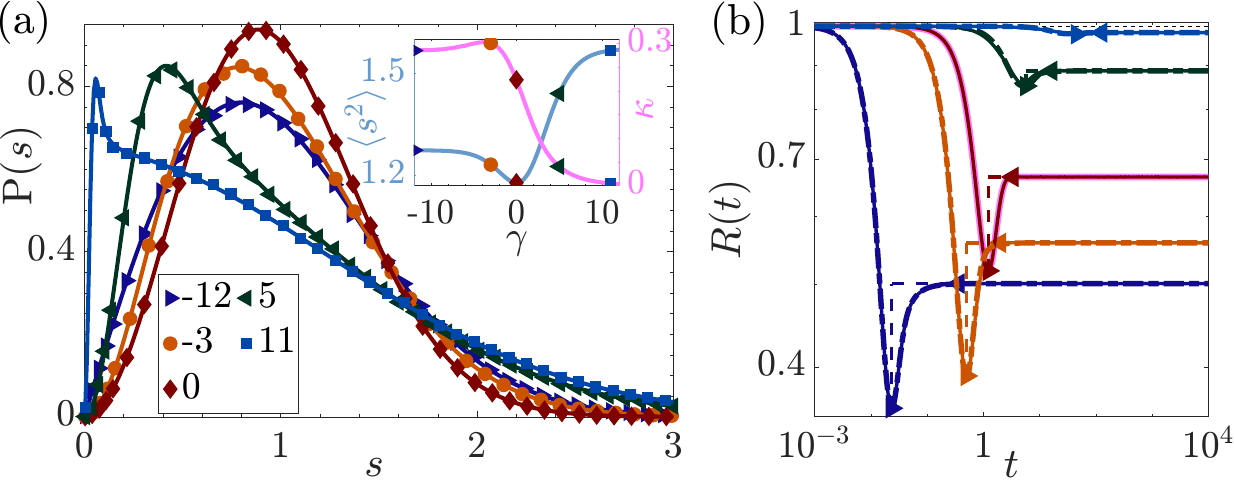}
	\caption{U-RPE: (a) density of NNS for various $\gamma$. Solid lines denote analytical form in equation~\ref{eq_s_URPE} wheres the markers denote simulation over $10^7$ samples. Inset shows $\mean{s^2}$, 2nd moment of NNS, where the blue solid line denotes analytical form in table~\ref{tab_s2} and $\kappa$, the relative depth of the correlation hole in the fidelity (pink solid line).
		(b) temporal evolution of fidelity where Thouless and relaxation times are denoted by markers and correlation hole is marked by dotted lines. Pink bold line is the analytical form for GUE (table~\ref{tab_WD_S_Rt}) and the dashed lines are fit using equation~\ref{eq_Rt_fit}. The asymptotic values are given in table~\ref{tab_R_bar}.
	}
	\label{fig_URPE}
\end{figure}
Then using equation~\ref{eq_S_def}, we get the density of NNS for $2\times 2$ U-RPE \cite{Das1}
\begin{equation}
	\label{eq_s_URPE}
	\pnns{2}{s} = \frac{2\mu^2}{\sqrt{\abs{1-2^\gamma}}}s e^{-\mu^2 s^2}g(\mu \sqrt{\abs{ 1 - 2^{-\gamma} }}s)
\end{equation}
where $\mu = \frac{2^{\frac{\gamma}{2}} + c_\gamma}{\sqrt{\pi}}$ and the functional forms of $c_\gamma$ and $g(x)$ are given in table~\ref{tab_s_URPE}. $\pnns{2}{s}$ for various $\gamma$ are shown in figure~\ref{fig_URPE}(a) where $\gamma = 0$ corresponds to the Gaussian Unitary Ensemble (GUE). Equation~\ref{eq_s_URPE} implies that $2\times 2$ U-RPE exhibits quadratic (linear) level repulsion for $\gamma\to 0$ ($\gamma\to -\infty$). We present the 2nd moment of NNS in table~\ref{tab_s2} and plot the same in the inset of figure~\ref{fig_URPE}(a) to show a crossover from linear to quadratic to no level repulsion.
\begin{table}[b]
	\centering
	\begin{tabular}{|c|c|c|c|}
		\hline
		& $c_\gamma$ & $g(s)$ & $f(s)$\\ \hline
		$\gamma<0$		& $\frac{ \cos^{-1} \del{2^{\frac{\gamma}{2}}} }{\sqrt{1 - 2^\gamma}}$ & $\ferf{s}$ & $\frac{\sqrt{\pi}}{2}\frac{(2s^2 - 1)\ferf{s}e^{s^2}}{s} + 1$\\ \hline
		$\gamma\geq 0$	& $\frac{ \cosh^{-1} \del{2^{\frac{\gamma}{2}}} }{\sqrt{2^\gamma - 1}}$ & $\ferfi{s}$ & $\frac{(2s^2 + 1)\fdsnf{s}}{s} - 1$ \\ \hline
	\end{tabular}
	\caption{U and S-RPE: $c_\gamma, g(s)$ for $\pnns{2}{s}$ (equation~\ref{eq_s_URPE}) and $f(s)$ for $\pnns{4}{s}$ (equation~\ref{eq_s_SRPE}).
	}
	\label{tab_s_URPE}
\end{table}
Having quantified the degree of level repulsion, here again we focus on the dynamical properties of $2\times 2$ U-RPE. We prepare the system in the initial state $\ket{\Psi_0} = \left( \begin{array}{c}
	1\\ 0
\end{array}\right)$. The eigenvectors $\ket{\Phi_{1, 2}}$ imply that the ensemble averaged fidelity of $\ket{\Psi_0}$ in a hermitian system is generically given by equation~\ref{eq_Rt_general8} and particularly for U-RPE, we obtain
\begin{equation}
	\label{eq_Rt_URPE}
	\eqalign{ &\mean{R(t)} = \overline{R} + \frac{e^{-t^2}}{2(2^{-\gamma} - 1)} + \frac{2^{\gamma-1}}{\sqrt{1 - 2^\gamma}}\int_{0}^{\infty} dS \cos\del{S t}f(S)\\
		&f(S) = e^{-2^{\gamma - 2}S^2}\ferf{\frac{\sqrt{1 - 2^\gamma}}{2} S}\del{\frac{S}{2} - \frac{1}{(1-2^\gamma) S}}
			}
\end{equation}
where $\overline{R}$ is given in table~\ref{tab_R_bar} and is trivially equal to the ensemble averaged IPR. In figure~\ref{fig_URPE}(b), we show the temporal evolution of fidelity for various $\gamma$. Equation~\ref{eq_Rt_URPE} implies that for very short time ($t\ll 1$), fidelity decays as $\mean{R(t)} = 1 - \frac{t^2}{2^\gamma}$ while $\kappa$ captures the crossover from chaotic to integrable regime, as shown in the inset of figure~\ref{fig_URPE}(a).

\section{Symplectic \rpe ~(S-RPE)}
 Now we study the symplectic ensemble to complete Dyson's threefold classification of the symmetry classes \cite{Dyson1} in case of RPE. A $2\times 2$ symplectic matrix is defined in terms of quaternions as
\begin{equation}
	\label{eq_H_sympl}
	\eqalign{ 	&H = \left(\begin{array}{cc}
			\hat{h}_{11}& \hat{h}_{12}\\
			\hat{h}_{21}& \hat{h}_{22}
		\end{array}\right),\quad \hat{h}_{mn} = \vec{h}_{mn} \cdot \vec{\tau}\\
				&\vec{h}_{mn} = \cbr{h_{mn}^{(0)}, h_{mn}^{(1)}, h_{mn}^{(2)}, h_{mn}^{(3)}},\quad \vec{\tau} = \cbr{\mathbb{I}, -i\sigma^x, -i\sigma^y, -i\sigma^z}
		}
\end{equation}
where $\sigma^{x,y,z}$ are the Pauli matrices and $\mathbb{I}$ is the identity matrix. $H$ has two unique eigenvalues with 2-fold Kramer's degeneracy \cite{Haake2}. As $H$ is hermitian, $h_{mn}^{(j)} = h_{nm}^{(j)}\in\mathbb{R}\;\forall\; j$, leading to six independent real elements.
\begin{figure}[t]
	\raggedleft
	\includegraphics[width=\x\textwidth]{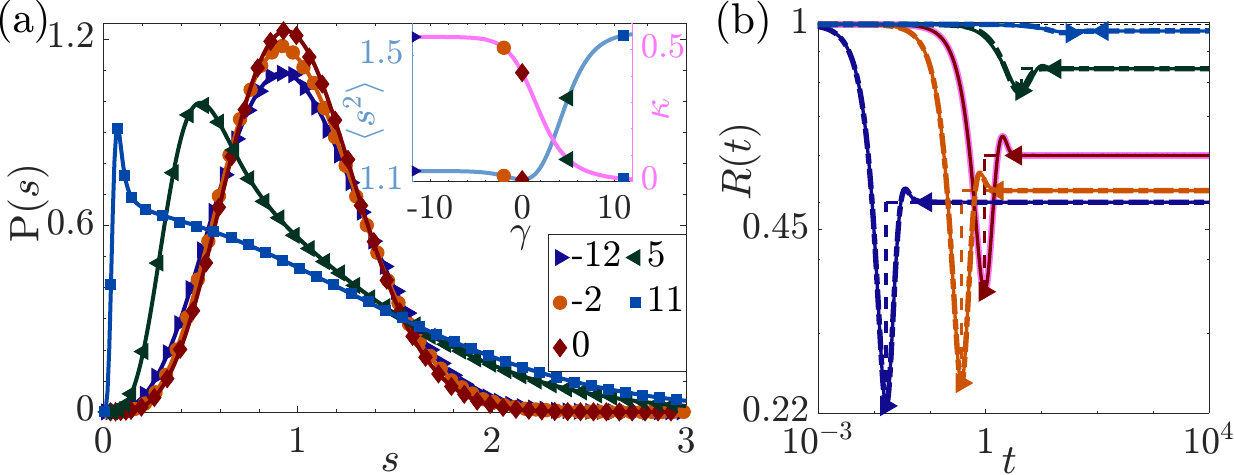}
	\caption{S-RPE: (a) density of NNS for various $\gamma$. Solid lines denote analytical form in equation~\ref{eq_s_SRPE} wheres the markers denote simulation over $10^7$ samples. Inset shows $\mean{s^2}$, 2nd moment of NNS, where the blue solid line denotes analytical form in table~\ref{tab_s2} and $\kappa$, the relative depth of the correlation hole in the fidelity (pink solid line).
		(b) temporal evolution of fidelity where Thouless and relaxation times are denoted by markers and correlation hole is marked by dotted lines. Pink bold line is the analytical form for GSE (table~\ref{tab_WD_S_Rt}) and the dashed lines are fit using equation~\ref{eq_Rt_fit}. The asymptotic values are given in table~\ref{tab_R_bar}.
	}
	\label{fig_SRPE}
\end{figure}
The eigenvectors of $H$ form a unitary matrix $U \equiv U(\zeta, \alpha, \theta, \phi)$ dependent on four independent elements. Then using the change of variables in table~\ref{tab_change}, we can express the joint density of non-degenerate energies and $\zeta$ in case of $2\times 2$ S-RPE
\begin{equation}
	\label{eq_P_E_theta_SRPE}
	\prob{E_1, E_2, \zeta} = \frac{4^{\gamma+1}}{2^9\pi} \abs{\sin\zeta}^3 \del{E_1-E_2}^4 e^{ - \frac{E_1^2 + E_2^2}{2} + \del{1-2^\gamma}\frac{(E_1-E_2)^2}{4}\sin^2\zeta }
\end{equation}
where $\zeta$ plays the role of $2\theta$ in case of O and U-RPE. For $\gamma = 0$, we retrieve the Gaussian Symplectic Ensemble (GSE). Integrating out $\zeta$ in equation~\ref{eq_P_E_theta_SRPE} yields the joint density of energy and using equation~\ref{eq_S_def}, we get the density of NNS for $2\times 2$ S-RPE
\begin{equation}
	\label{eq_s_SRPE}
	\eqalign{ &\pnns{4}{s} = \frac{2^{2\gamma+1}\mu^3}{\sqrt{\pi |1-2^\gamma|^5}} s^2\exp\del{-\frac{\mu^2 s^2}{|1-2^\gamma|}} f(\mu s)\\
		      &\mu = \frac{1}{ \sqrt{\pi\abs{1 - 2^\gamma}} }\abs{ \frac{3}{2} - 2^\gamma - 2^{\frac{\gamma}{2}} \del{2-\frac{3}{2^{\gamma+1}}} c_\gamma }
		    }
\end{equation}
where $f(x)$ and $c_\gamma$ are given in table~\ref{tab_s_URPE}. $\pnns{4}{s}$ exhibits cubic ($\gamma\to-\infty$), quartic ($\gamma = 0$) or no ($\gamma\to\infty$) level repulsion depending on $\gamma$. Note that in $2\times 2$ RPE from all the three symmetry classes, the degree of level repulsion for $\gamma\to-\infty$ is precisely one less than that in the Wigner-Dyson (i.e.~$\gamma = 0$) limit. We show the 2nd moment of NNS in table~\ref{tab_s2} and plot the same in the inset of figure~\ref{fig_SRPE}(a) to show a crossover among various degrees of level repulsion.

Proceeding as in the previous sections, average fidelity of $\ket{\Psi_0}$ for $2\times 2$ S-RPE becomes
\begin{equation}
	\label{eq_Rt_SRPE}
	\eqalign{ 	&\mean{R(t)} = \overline{R} + \frac{2^{2\gamma - 3}}{\sqrt{\pi}}\int_{0}^{\infty} dS \cos\del{S t}f(S)\\
				&f(S) = \frac{ (2^\gamma - 1)s^3 + 4s + \frac{12}{s(2^\gamma - 1)}}{\del{2^\gamma - 1}^{\frac{3}{2}}} - \frac{s^2 + \frac{6}{2^\gamma - 1}}{2^\gamma - 1}
			}
\end{equation}
where $\overline{R}$ is given in table~\ref{tab_R_bar}. Here also $\kappa$ captures the crossover from chaotic to integrable regime, as shown in the inset of figure~\ref{fig_SRPE}(a). Importantly, a $2\times 2$ self-dual matrix from S-RPE has two-fold Kramers' degeneracy and dimension 4 in the computational basis, hence the ensemble averaged IPR does not coincide with $\overline{R}$.

\begin{table}[b]
	\raggedleft
	\begin{tabular}{|c|c|c|}
		\hline
		Model		& $\prob{s}$												& $\mean{R(t)}$ \\
		\hline
		Poisson		& $\frac{2}{\pi}e^{-\frac{s^2}{\pi}}$					& $1$ \\
		\hline
		GOE			& $\frac{\pi}{2}se^{-\frac{\pi}{4}s^2}$				& $1 - \frac{1}{2}t\fdsnf{t}$ \\
		\hline
		GUE			& $\frac{32}{\pi^2}s^2e^{-\frac{4}{\pi}s^2}$  		& $\frac{2}{3} + \frac{1}{3}\del{1-2t^2}e^{-t^2}$ \\
		\hline
		GSE			& $\frac{2^{18}}{3^6\pi^3}s^4e^{-\frac{64}{9\pi}s^2}$	& $\frac{3}{5} + \frac{1}{15}\del{6 + 8t^2\del{t^2 - 3}}e^{-t^2}$ \\
		\hline
		\bte		& $a_\beta s^\beta e^{-b_\beta^2 s^2}$				& $\overline{R} + \del{1 - \overline{R}} {}_1F_1\del{\frac{\beta+1}{2}; \frac{1}{2}; -t^2}$ \\
		\hline
	\end{tabular}
	\caption{Poisson, Wigner-Dyson and \bte: Density of NNS and mean fidelity, where $\fdsnf{x} = e^{-x^2}\int_{0}^{x}e^{y^2}dy = \frac{\sqrt{\pi}}{2}e^{-x^2}\ferfi{x}$ is the Dawson integral. For \bte, $a_\beta = \frac{ 2\fgamma{1 + \frac{\beta}{2}}^{\beta+1} }{ \fgamma{\frac{\beta+1}{2}}^{\beta+2} }$ and $b_\beta = \frac{ \fgamma{1 + \frac{\beta}{2}} }{ \fgamma{\frac{\beta+1}{2}} }$ in the density of NNS and $\overline{R} = 1 - \frac{\beta}{2\del{\beta+1}}$ in the average fidelity. Note that \bte\ encompasses the results for Poisson and WDE for respective values of $\beta$.
	}
	\label{tab_WD_S_Rt}
\end{table}

\section{$\beta$-Rosenzweig-Porter Ensemble ~(\brpe)}
So far we have studied the density of NNS and the average fidelity in case of RPE from three symmetry classes consisting of real, complex and quaternion matrices where the Dyson's index, $\beta$ can be only 1, 2 and 4, respectively and $\gamma$ controls the disorder strength. Particularly for $\gamma = 0$, we get closed form expressions for the $\mean{R(t)}$ of GOE, GUE, GSE (table~\ref{tab_WD_S_Rt}). Now we allow $\beta$ to be any real number such that the joint density of eigenvalues follows the \bte\ \cite{Forrester3, Marino1}. Various canonically invariant matrix models of \bte\ are proposed till date \cite{Vivo1, Allez4, Cunden1}. However, a convenient matrix representation of \bte\ can be found at the cost of canonical invariance, where the relevant ensemble  consists of random real symmetric tridiagonal matrices \cite{Dumitriu1}. Such a matrix model hosts three distinct phases separated by two 2nd order critical points \cite{Das2, Das5}. We can further generalize \bte\ if we let the relative strength of diagonal and off-diagonal elements to vary. In a simple two-level set-up, we introduce a two-parameter ensemble of $2 \times 2$ real matrices, the $\beta$-Rosenzweig-Porter ensemble (\brpe):
\begin{equation}
	\label{eq_brpe_H}
	H = \left(\begin{array}{cc}
			x & z\\
			z& y
		\end{array}\right),\: x, y\sim \mathcal{N}\del{0, 1},\: \frac{z}{\sigma} \sim \chi_\beta,\: \sigma^2 = \frac{1}{2^{\gamma+1}}
\end{equation}
where $\chi_\beta$ is the chi distribution with degree of freedom $\beta$ and $x, y, z$ are mutually independent. Thus in \brpe, $\gamma$ controls the relative strength of the disorder and $\beta$ is the fictitious inverse temperature of the analogous Coulomb gas model \cite{Dyson2}. We will show that \brpe\ encompasses all the results given for previous RPEs. Equation~\ref{eq_brpe_H} implies that the density in matrix space is
\begin{equation}
	\label{eq_P_H_bRPE}
	\prob{H} = \frac{1}{\sigma^\beta 2^\frac{\beta}{2} \pi \fgamma{\frac{\beta}{2}}} z^{\beta-1} \exp\del{-\frac{x^2 + y^2 + \frac{z^2}{\sigma^2}}{2}}
\end{equation}
Since $H$ in equation~\ref{eq_brpe_H} is a real symmetric matrix, corresponding eigenvectors parameterized by $\theta$ forms an orthogonal (rotation) matrix and $H$ can be expressed in terms of the energies, $E_{1, 2}$ and $\theta$ as in equation~\ref{eq_H_O-RPE}. However, the off-diagonal element $z$ is always positive, hence $\theta\in[0, \frac{\pi}{2}]$ for $E_1\geq E_2$ and $\theta\in(\frac{\pi}{2}, \pi]$ for $E_1 < E_2$. Then we can transform the density in matrix space (equation~\ref{eq_P_H_bRPE}) to the density in eigenspace
\begin{equation}
	\label{eq_P_E_theta_bRPE}
	\prob{E_1,E_2,\theta} = \frac{ \abs{E_1-E_2}^\beta \abs{\sin 2\theta}^{\beta-1} }{\sigma^\beta 2^{\frac{3\beta}{2}} \pi \fgamma{\frac{\beta}{2}}} e^{ - \frac{E_1^2+E_2^2}{2} + \del{1 - \frac{1}{2\sigma^2}}\frac{\sin^2 2\theta}{4} \del{E_1-E_2}^2 }.
\end{equation}
Note that other than the normalization constant, $\prob{E_1,E_2,\theta}$ matches with those of O, U and S-RPE (equation~\ref{eq_P_E_theta_ORPE}, \ref{eq_P_E_theta_URPE} and \ref{eq_P_E_theta_SRPE}) for respective values of $\beta$. Integrating out $\theta$ in equation~\ref{eq_P_E_theta_bRPE} gives the joint density of energies, allowing us to compute the density of NNS
\begin{equation}
	\label{eq_s_bRPE}
	\eqalign{ 	&\pnns{\beta}{s} = \mu f(\mu s)\\
				&f(s) = \frac{2^{\frac{\beta}{2}(\gamma - 2)}}{ \fgamma{\frac{\beta+1}{2}} } S^\beta e^{ - \frac{S^2}{4} } {}_1F_1\del{\frac{\beta}{2}; \frac{\beta+1}{2}; \del{1 - 2^\gamma}\frac{S^2}{4}}\\
				&\mu = \frac{ 2^{\frac{\beta\gamma}{2}}\beta \fgamma{\frac{\beta}{2}} }{ \fgamma{\frac{\beta+1}{2}} } {}_2F_1\del{\frac{\beta}{2}, \frac{\beta+2}{2}; \frac{\beta+1}{2}; 1-2^\gamma}
			}
\end{equation}
where ${}_1F_1(a; b; z)$ is the Kummer confluent hypergeometric function and ${}_2F_1(a, b; c; z)$ is the hypergeometric function. $\pnns{\beta}{s}$ converges to the density of NNS for O, U and S-RPE (equation~\ref{eq_s_ORPE}, \ref{eq_s_URPE}, \ref{eq_s_SRPE}) for $\beta = 1, 2$ and 4, respectively, while for $\gamma = 0$, we get the generalized Gamma distribution valid for $2\times 2$ \bte\ \cite{Caer1}. The 2nd moment of NNS for \brpe\ is given in table~\ref{tab_s2}.
\begin{figure}[t]
	\centering
	\includegraphics[width=0.36\textwidth]{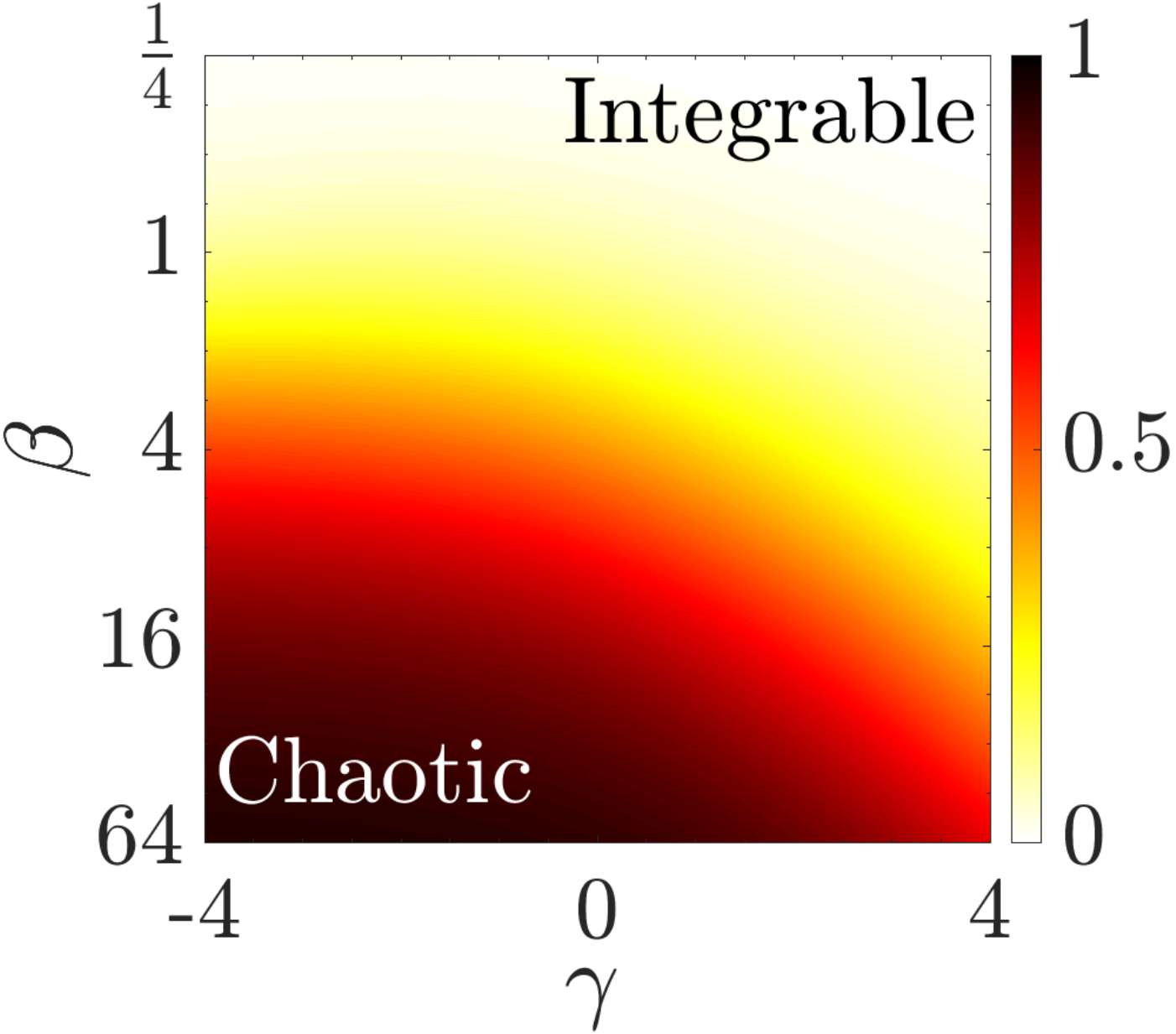}
	\caption{\brpe: $\kappa$, relative depth of the correlation hole in the $\gamma$-$\beta$ plane showing the crossover between chaotic and integrable regime.}
	\label{fig_col_bRPE}
\end{figure}

To understand the unitary time evolution of the state $\ket{\Psi_0} = \left(\begin{array}{c}
	1\\ 0
\end{array}\right)$, we apply equation~\ref{eq_Rt_general} to equation~\ref{eq_P_E_theta_bRPE} to get the ensemble averaged fidelity
\begin{equation}
	\label{eq_Rt_bRPE}
	\eqalign{ 	&\mean{R(t)} = \overline{R} + \frac{ \beta 2^{\frac{\beta\gamma}{2} - 3} }{ \fgamma{\frac{\beta+3}{2}} } \int_{0}^{\infty} dS\; \cos\del{St} f(S)\\
				&f(S) = S^\beta e^{-\frac{S^2}{4}} {}_1F_1\del{\frac{\beta+2}{2}; \frac{\beta+3}{2}; \del{1 - 2^\gamma}\frac{S^2}{4}}
			}
\end{equation}
where $\overline{R}$ is given in table~\ref{tab_R_bar}. Equation~\ref{eq_Rt_bRPE} yields the fidelity for O, U and S-RPE (equation~\ref{eq_Rt_ORPE}, \ref{eq_Rt_URPE}, \ref{eq_Rt_SRPE}) for respective values of $\beta$. For very short time, average fidelity of \brpe\ decays as $\mean{R(t)} \sim 1 -\beta\sigma^2 t^2$. 
Since the relative depth of the correlation hole, $\kappa$, depends on both energy and state structure, we show $\kappa$ in the $\gamma$-$\beta$ plane in figure~\ref{fig_col_bRPE} to identify a crossover between chaotic and integrable regimes.

For $\gamma = 0$, we obtain a closed form valid for $2\times 2$ \bte\
\begin{equation}
	\label{eq_Rt_beta}
	\mean{R(t)} = \overline{R} + \del{1 - \overline{R}} {}_1F_1\del{\frac{\beta+1}{2}; \frac{1}{2}; -t^2}, \quad \overline{R} = 1 - \frac{\beta}{2\del{\beta+1}}
\end{equation}
We observe that $\beta$, the degree of level repulsion, appears as the only parameter in equation~\ref{eq_Rt_beta}. Hence for any $2\times 2$ matrix from any symmetry class, we propose a two parameter ansatz for the fidelity of the initial state $\ket{\Psi_0}$ to follow
\begin{equation}
	\label{eq_Rt_fit}
	\frac{R(a, b; t) - \overline{R}}{1 - \overline{R}} = {}_1F_1\del{\frac{b+1}{2}; \frac{1}{2}; -a^2t^2}
\end{equation}
where $a$ scales the time axis and $b$ controls the rate of initial decay of the fidelity. In figures~\ref{fig_ORPE}, \ref{fig_URPE} and \ref{fig_SRPE}(b), we show that equation~\ref{eq_Rt_fit} provides a good fit for the fidelity for all values of $\gamma$ across any symmetry class. Thus similar to Brody distribution serving as an empirical form of density of NNS, our ansatz in equation~\ref{eq_Rt_fit} works as a fitting function for fidelity of any 2-level system.
\section{\rpe\ for large $N$}
\begin{figure}[t]
	\raggedleft
	\includegraphics[width=\textwidth]{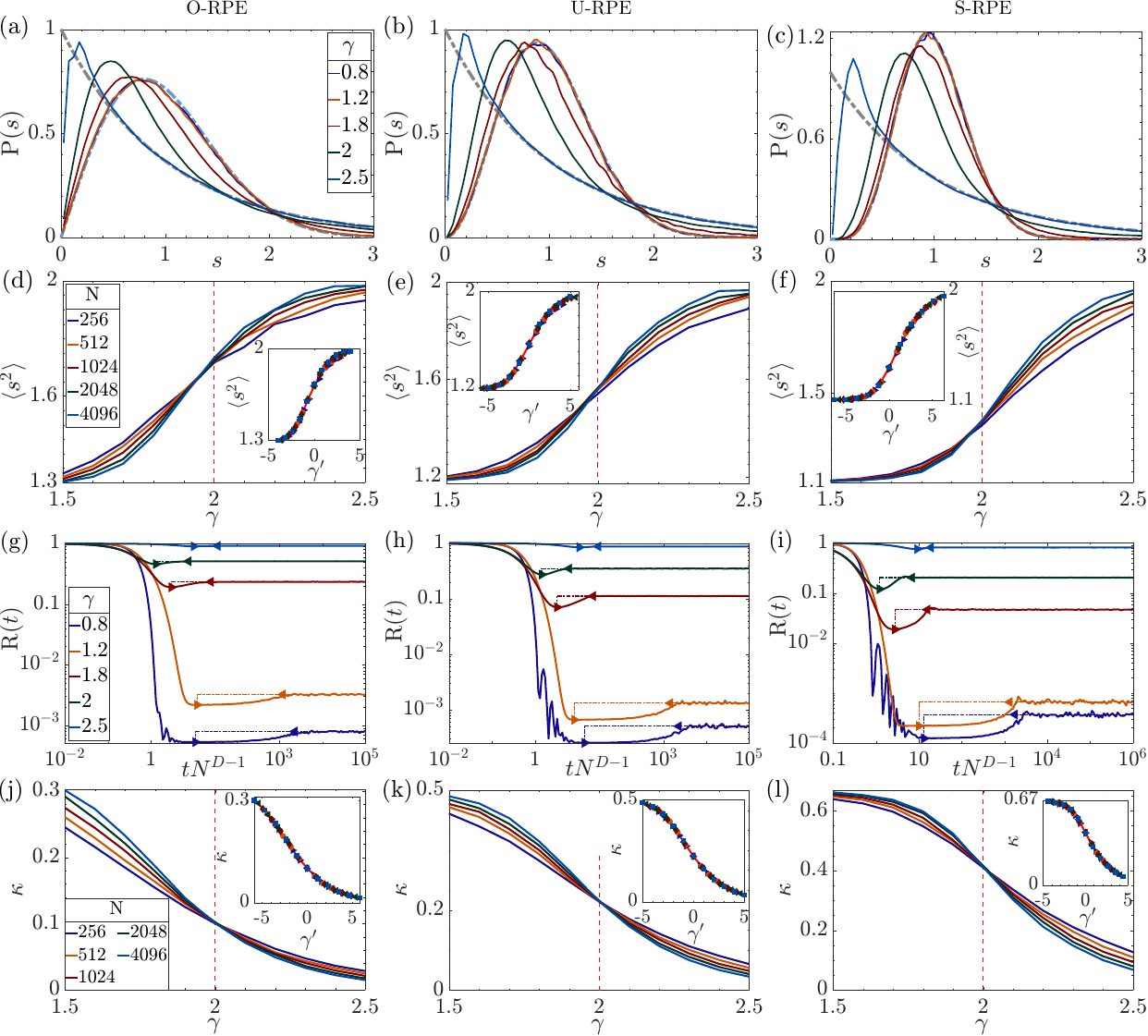}
	\caption{RPE for large $N$: 1st, 2nd and 3rd columns show the results for orthogonal, unitary and symplectic \rpe, respectively. Densities of NNS of the unfolded eigenvalues (obtained from middle 25\% of the spectrum) are shown for varying $\gamma$ values in (a), (b) and (c) ($N=4096$). Dashed lines denote the analytical curves for $2\times 2$ WDE and Poisson ensemble. In (d), (e) and (f), the 2nd moment of NNS is shown for varying $N$. In (g), (h) and (i), we show the time evolution of the fidelity of $\frac{N}{4}$ $\delta$-localized states with energy closest to 0 for various $\gamma$ and $N = 4096$. Dashed lines denote the equilibrium values and markers denote the Thouless and relaxation times. In (j), (k) and (l), $\kappa$, the relative depth of the correlation hole is shown for varying $N$. The insets of (d), (e), (f), (j), (k) and (l) show the collapsed data using 2nd order phase transition ansatz, $\gamma' = (\gamma - 2)(\log N)^\frac{1}{\nu}$ \cite{Das2}. The estimated critical exponent $\nu \approx 1.2025, 0.988,  0.9765$ for $\mean{s^2}$ and $\nu \approx 1.0102, 1.0745, 1.1414$ for $\kappa$ from the three symmetry classes, respectively. The data are averaged over 128 disorder realizations.
	}
	\label{fig_RPE_N}
\end{figure}
In this section, we numerically compute the statistical and dynamical  properties of RPE for large matrix dimensions, $N$
\begin{equation}
	\label{eq_H_large_N_element}
	\mean{H_{n, n}^2} = 1,\quad \mean{\abs{H_{n, m}}^2} = \frac{\beta}{2N^\gamma},\quad \gamma\in\mathbb{R},
\end{equation}
and compare them with those of $2\times 2$ systems obtained earlier. 
The density of NNS for $2\times 2$ WDE (RPE with $\gamma = 0$) is a good approximation for that of the larger matrices from the same ensemble \cite{Haake2}, consequently 2nd moment of NNS for $N = 2$ and large $N$ are close to each other (table~\ref{tab_s2}). However, this is not true for the RPE with $\gamma\neq 0$ since in $2\times 2$ RPE, the density of NNS interpolates between clustering $\prob{s}\sim e^{-s^2}$ and repulsion $\prob{s}\sim s^\beta e^{-s^2}$ whereas for $N\gg 1$, the clustering is of the form $\prob{s} = e^{-s}$ in the integrable limit. 
 In figure~\ref{fig_RPE_N}(a), (b) and (c), we show the densities of NNS of the unfolded eigenvalues (from the middle 25\% of the spectrum averaged over 128 disordered realizations for $N = 4096$) for various $\gamma$ values along with the analytical expressions of the $2\times 2$ WDE and Poisson ensemble. To identify the transition from level repulsion to clustering, we show the 2nd moment of NNS as a function of $\gamma$ for various $N$ in figure~\ref{fig_RPE_N}(d), (e) and (f). The crossover curves get steeper with $N$ and tend to intersect each other around $\gamma = 2$. Using the 2nd order phase transition ansatz \cite{Das2}, we are able to collapse the $\mean{s^2}$ data from all system sizes, as shown in the respective insets. Therefore, short-range energy correlations for RPE from all the three symmetry classes show a 2nd order phase transition at $\gamma = 2$, which was previously established using ratio of level spacing \cite{Pino1}.

Having shown the statistical measure of chaos-integrability transition in RPE, we now look at the corresponding dynamical signatures. For $\ket{\Psi}$, the $\delta$-localized states with energy close to 0 (i.e.~$\bra{\Psi}H\ket{\Psi}\sim 0$), we compute the ensemble averaged time evolution of the fidelity
\begin{equation}
	\label{eq_Rt_gen}
	\mean{R(t)} = \abs{\braket{\Psi|\Psi(t)}}^2 = \sum_{m, n} |c_m|^2|c_n|^2 e^{i(E_m - E_n)t}
\end{equation}
where $c_n = \braket{E_n|\Psi}$ is the $n$th component of $\ket{\Psi}$ in the energy eigenbasis $\{ \ket{E_n}\}$. Equation~\ref{eq_Rt_gen} implies that the asymptotic value of fidelity is $\overline{R} = \sum_n |c_n|^4$, which is the IPR of the initial state in $\{ \ket{E_n}\}$, provided the energy spectrum is non-degenerate \cite{Schiulaz1}. However, self-dual quaternion matrices from S-RPE show two-fold Kramers' degeneracy, hence the asymptotic value of fidelity becomes
\begin{equation}
	\label{eq_R_bar_S}
	\overline{R}_\mathrm{S-RPE} = \sum_n |c_n|^4 + 2 \sum_{m = 1, 3, 5,\dots} |c_m c_{m+1}|^2.
\end{equation} 
Therefore, degeneracy in the energy spectrum increases the asymptotic value of fidelity, i.e.~an initially localized excitation spreads over larger fraction of the total Hilbert space volume at long time. In figure~\ref{fig_RPE_N}(g), (h) and (i), we show the time evolution of the ensemble averaged fidelity for RPE from all the three symmetry classes for $N = 4096$ and various values of $\gamma$. We also mark the Thouless and relaxation times in each case and show the asymptotic value of fidelity via dashed lines. Note that, the Thouless time increases with system size as $\tTh\propto N^{1-D}$ where $D$ is the typical fractal dimension of the bulk eigenstates \cite{Tomasi1}. Hence, in the non-ergodic ($1<\gamma<2$, $D = 2-\gamma$) and localized ($\gamma\geq 2$, $D = 0$) regimes, the Thouless time increases exponentially with system size and we scale the time axis as $t\to tN^{D-1}$. The relative depths of the correlation hole are shown in figure~\ref{fig_RPE_N}(j), (k) and (l) as a function of $\gamma$ for various $N$. Corresponding insets show the collapsed data implying a 2nd order phase transition from chaos to integrability at $\gamma = 2$ for RPE from all the three symmetry classes. Therefore, $\kappa$ is a robust measure of dynamical signature of chaos-integrability crossover, which is particularly important for experiments with lack of knowledge of the energy levels \cite{Torres4}.

\section{Conclusions}
In this work, we study the energy correlations and the dynamical properties of $2\times 2$ generalized random matrix ensembles. In each of the Dyson's threefold classifications, i.e.~orthogonal, unitary and symplectic symmetry classes, we derive an analytical expression of the density and 2nd moment of NNS from the eigenvalue statistics for $2\times 2$ RPE. As the disorder strength ($\gamma$) is varied, the 2nd moment (table~\ref{tab_s2}) provides an exact quantification of the crossover among various degrees of level repulsion providing a better method than any empirical fitting. 
A general matrix model, \brpe, has been introduced, where in addition to the disorder strength, the Dyson index, $\beta$, can be varied continuously such that \brpe\ reproduces the results of the above symmetry classes for $\beta$ = 1, 2 and 4. In such a generalized matrix model, our aim has been to determine the crossover in the two-parameter ($\gamma, \beta$) phase plane in terms of both spectral and dynamical properties.

In order to study the dynamics, we prepare our two-level systems in the eigenstate of an integrable Hamiltonian and look at the time evolution of the corresponding fidelity. Any non-degenerate 2-level system undergoes Rabi oscillation if not prepared in its eigenstate \cite{Merlin1}. However, ensemble averaging suppresses the Rabi oscillation, and the average fidelity, $\mean{R(t)}$ relaxes to an equilibrium value $\overline{R}$, as ensemble averaging for a fixed initial state is equivalent to a mixed-unitary quantum channel destroying the coherence in the energy eigenbasis \cite{Roubeas1}. For the general \brpe, we analytically obtain the temporal evolution of $\mean{R(t)}$ which reduces to simpler expressions (table~\ref{tab_WD_S_Rt}) for the above-mentioned symmetry classes.
 We observe that upon increasing the system parameter $\gamma$ (therefore reducing the relative strength of hopping disorder) in any RPE, ensemble averaged fidelity takes longer time to equilibrate while the equilibrium value, $\overline{R}$, approaches unity. In general, $\mean{R(t)}$, after an initial quadratic decay, exhibits a minimum at Thouless time and equilibrates to $\overline{R}$ with a characteristic relaxation time-scale.
 We propose an ansatz (equation~\ref{eq_Rt_fit}) for the average fidelity of any 2-level system, therefore providing an empirical expression characterizing the dynamical crossover from chaos to integrability. We numerically estimate the relative depth of the correlation hole, which quantifies the crossover in qualitative agreement with that observed in the second moment of NNS. 
 Such a crossover in \brpe ~can be physically understood as the manifestation of the deformation of the symmetry present in an integrable Hamiltonian. In spite of the analytical simplicity of $2\times 2$ \brpe, corresponding $\mean{R(t)}$ captures the generic dynamical feature observed in large dimensional random matrices and realistic spin Hamiltonians \cite{Torres1, Torres6, Schiulaz1, Torres4, Torres5}. 
{We have also numerically computed the 2nd moment of NNS and the fidelity for large $N$ and obtain qualitatively identical features as shown for $2 \times 2$ RPE.  We find that both the 2nd moment of NNS and the relative depth of the correlation hole exhibit a second order phase transition at $\gamma=2$.}
 
Our results can be used to understand the fidelity of information transfer in a pre-engineered quantum wire. Note that a $d$-dimensional Cartesian product of a linear chain of two qubits governed by the Pauli matrix $\hat{\sigma}_x$ in the first excitation subspace allows perfect transfer of excitation between the antipodal points of the resulting hypercube geometry for any $d$ \cite{Christandl1, Christandl2}. Therefore, a Cartesian product of two-qubit chains governed by RPE may present a novel mechanism for controlling the transfer fidelity. Another interesting problem will be to consider a dimer governed by an electronic Hamiltonian from $2\times 2$ RPE and place it in a monochromatic quantized electromagnetic cavity to achieve a generalization of the Jaynes-Cummings model \cite{Larson1}, where controlling the extent of integrability of the dimer Hamiltonian may affect the population dynamics in a non-trivial manner. Therefore, our dynamical characterization of the loss of integrability in generalized matrix ensembles can have many potential applications.

\paragraph*{\bf Acknowledgment:} AKD is supported by an INSPIRE Fellowship, DST, India and the Fulbright-Nehru grant no. 2879/FNDR/2023-2024.

\paragraph*{\bf Author contribution statement:} AKD and AMG performed the analysis and wrote the paper.
\newpage
\appendix
\setcounter{figure}{0}
\setcounter{table}{0}
\section{Global scaling vs.~unfolding}\label{sec_A_unfold}
To obtain a universal form of the energy correlations for $N\gg 1$, the eigenvalues need to be unfolded to get rid of the global shape of the density of states \cite{Guhr1}. However for $N = 2$, mean level spacing has the same order of magnitude as the global bandwidth, hence the eigenvalues fluctuate around their mean positions in the same length scale as the global bandwidth. Therefore, unfolding leads to loss of universality for two-level systems. For example, unfolding in case of the Poisson ensemble yields $\prob{s}_{N=2} = 1 - \frac{s}{2},\: s\in[0, 2]$ whereas $\prob{s} = e^{-s}$ in the thermodynamic limit ($N\to\infty$). Hence for 2-level systems, we use the global scaling $s = \frac{S}{\mean{S}}$ to enforce $\int_{0}^{\infty}ds\; \prob{s} = 1$ (normalization) and $\int_{0}^{\infty}ds\; s\prob{s} = 1$ (i.e.~$\mean{s} = 1$). This ensures that Wigner's surmise is obtained for $2\times 2$ GOE whereas unfolding produces a significantly different from \cite{Berry1}.
\section{Fidelity of a general state for O-RPE}\label{sec_A_gen_Rt}
\begin{figure}[b]
	\raggedleft
	\includegraphics[width=\textwidth]{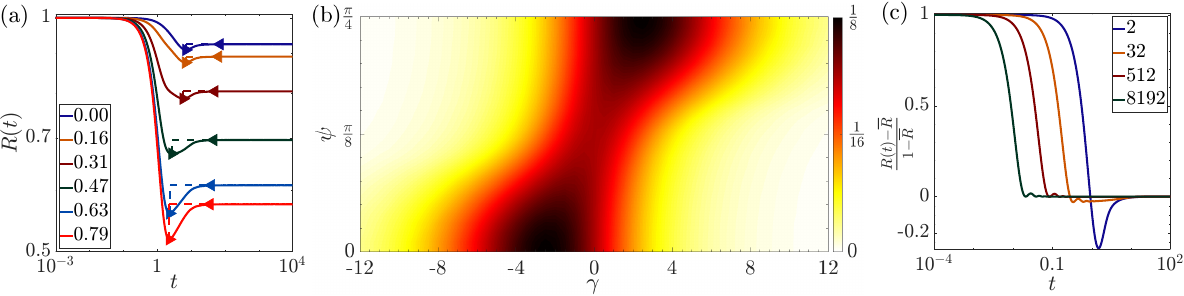}
	\caption{(a) Ensemble averaged fidelity for $2\times 2$ O-RPE for $\gamma = 5$ and different initial states parameterized by $\psi$ (equation~\ref{eq_Rt_gen_ORPE}). Thouless and relaxation times are shown via marker and the correlation hole is marked by dashed lines.
		(b) $\kappa$, the relative depth of correlation hole for $2\times 2$ O-RPE in $\gamma$-$\psi$ plane.
		(c) GOE, scaled fidelity for various $N>2$ \cite{Torres5} and $N = 2$ (Table~\ref{tab_WD_S_Rt}). 
	}
	\label{fig_A_1}
\end{figure}
Recall that the eigenvectors of a real symmetric matrix constitute a rotation matrix parameterized by $\theta$. Then for an initial state, $\ket{\Psi_0} = \left(\begin{array}{c}
	\cos\psi\\ \sin\psi
\end{array}\right)$, the fidelity becomes $R(t) = 1 - \sin^2\del{2\theta + 2\psi}\sin^2\frac{St}{2}$ where $S = |E_2-E_1|$. Then for $2\times 2$ O-RPE, ensemble averaged fidelity is
\begin{equation}
	\label{eq_Rt_gen_ORPE}
	\eqalign{ &\mean{R(t)} = \overline{R} + 2^{\frac{\gamma}{2}-3}\int_{0}^{\infty} dS\; \cos\del{St} f(S)\\
		&f(S) = S e^{ \del{c_\gamma - \frac{1}{4}} S^2 } \sbr{ \fbsli{0}{c_\gamma S^2} + \fbsli{1}{c_\gamma S^2}\cos 4\psi }\\
		&\overline{R} = \frac{3}{4} + \frac{\cos 4\psi}{4}\del{1 - \frac{2}{1 + 2^{\frac{\gamma}{2}}}}
	}
\end{equation}
which is a slightly modified version of equation~\ref{eq_Rt_ORPE}. Equation~\ref{eq_Rt_gen_ORPE} implies that $\mean{R(t)}$ is independent of the initial state for $\gamma = 0$ (i.e.~GOE limit) which can be attributed to the orthogonal invariance present at $\gamma = 0$. Due to $4\psi$ appearing in the argument of cosine function, we only need to look at $0\leq \psi\leq \frac{\pi}{4}$ where $\ket{\Psi_0}$ changes from $\left(\begin{array}{c}
	1\\ 0
\end{array}\right)$ to $\frac{1}{\sqrt{2}}\left(\begin{array}{c}
    1\\ 1
\end{array}\right)$. In figure~\ref{fig_A_1}(a), we show the mean fidelity for various initial states for $\gamma = 5$ while $\kappa$, the relative depth of the correlation hole is shown in figure~\ref{fig_A_1}(b). However, $\kappa$ is a measure of integrability only for $\psi = 0$ where $\ket{\Psi_0}$ is an eigenstate of the integrable systems.

\section{RPE as deformed ensemble}\label{sec_A_deformed}
\begin{table}[b]
	\begin{tabular}{|c|c|c|}
		\hline
		$\ot$ 		& 	$H_+$		& $H_-$
		\\ \hline
		$\left(\begin{array}{cc}
			\cos\xi& \sin\xi\\
			\sin\xi& -\cos\xi
		\end{array}\right)$ & $\left(\begin{array}{cc}
			x& \frac{\tan\xi}{2}(x-z)\\
			\frac{\tan\xi}{2}(x-z)& z
		\end{array}\right)$ & $y\left(\begin{array}{cc}
			-\tan\xi& 1\\
			1& \tan\xi
		\end{array}\right)$ 
		\\ \hline
		$\left(\begin{array}{cc}
			\cos\xi& \sin\xi e^{-i\alpha}\\
			\sin\xi e^{i\alpha}& -\cos\xi
		\end{array}\right)$ & $\left(\begin{array}{cc}
			x& \frac{\tan\xi}{2}(x-z)e^{-i\alpha}\\
			\frac{\tan\xi}{2}(x-z)e^{i\alpha}& z
		\end{array}\right)$ & $\left(\begin{array}{cc}
			-u\cos\phi'\tan\xi& v+iw\\
			v-iw& u\cos\phi'\tan\xi
		\end{array}\right)$
		\\ \hline
	\end{tabular}
	\caption{Structures of the symmetry operator $\ot$, integrable Hamiltonian, $H_+$ and the symmetry breaking Hamiltonian $H_-$ where first and second row corresponds to symmetric and hermitian matrices, respectively. In the diagonalizing basis of $\ot$, we can reduce the symmetric matrices as: $H_+\to \frac{x+z}{2}\mathbb{I} + \frac{x-z}{2}\sigma_z$ and $H_-\to y\sec\xi \sigma_x$. For hermitian $H_-$, $u = \sqrt{v^2 +w^2}, \phi' = \alpha + \phi$ and $\phi = \tan^{-1}\frac{w}{v}$.
	}
	\label{tab_deformed}
\end{table}
In RPE, all the symmetries present in an integrable Hamiltonian is broken in a mean-field manner. Hence RPE can be thought as a deformed ensemble. This can be easily illustrated in case of $N = 2$. Any $2\times 2$ matrix from O-RPE can be written as
\begin{equation}
	H = H_+ + 2^{-\frac{\gamma}{2}}H_-,\quad H_+ = y\mathbb{I} + z\sigma_z,\: H_- = x\sigma_x
\end{equation}
where $x, y, z$ are mutually independent and follow $\mathcal{N}\del{0, \frac{1}{2}}$, i.e.~Gaussian distribution with mean 0 and variance $\frac{1}{2}$. Since $\cbr{\sigma_x, \sigma_z} = 0$, the symmetry present in $H_+$ is deformed by $H_-$ where $\gamma$ controls the perturbation strength. Note that the perturbation strength diverges for $\gamma\to -\infty$, leading to inflation of eigenvalues as in the Pechukas-Yukawa gas \cite{Steeb1}. Therefore we can regularize the O-RPE following \cite{Haake3}
\begin{equation}
	\label{eq_H_deformed}
	\eqalign{ &H = \cos\alpha H_+ + \sin\alpha H_-\\
		\Rightarrow &\prob{S, \theta} = \frac{S}{\pi \sin 2\alpha} e^{-\del{\frac{\sin^2 2\theta\cot 2\alpha}{\sin 2\alpha} + \frac{1}{4\cos^2\alpha}}S^2}
	}
\end{equation}
where $S$ is the energy gap and $\theta$ parameterizes the eigenstates. Equation~\ref{eq_H_deformed} ensures that the eigenvalues are always bounded.

Alternatively, starting from a general symmetry operator, $\ot$, we can obtain the integrable Hamiltonian and its symmetry breaking counterpart (table~\ref{tab_deformed}). Then we can define $H(\lambda) = H_+ + \lambda H_-$, where $\lambda$ controls the perturbation strength. $H(\lambda)$ can be mapped to RPE as we identify the perturbation strength as the variance of off-diagonal elements in RPE. Therefore we can establish a correspondence between symmetry breaking and disorder in $2\times 2$ RPE.

\section{Comparison of fidelity between $N = 2$ and large $N$}\label{sec_A_Rt_compare}
In case of GOE, the analytical expression of fidelity for $N\gg 1$ \cite{Torres5} is compared with $\mean{R(t)}_{N = 2}$ in table~\ref{tab_Rt_GOE_comparison}. We observe that $\mean{R(t)}$ behaves in a qualitatively similar way for both $N = 2$ and $N \gg 1$ with the exception of an oscillation decaying as $\propto t^{-3}$ in the intermediate time for $N\gg 1$. Similarly for GUE, $\mean{R(t)}$ attains minimum at $\tTh = \sqrt{\frac{3}{2}}\approx 1.22474$ where $\mean{R(\tTh)} = \frac{2}{3}\del{1 - e^{-\frac{3}{2}} }\approx 0.517913$. On the other hand, $\tR\approx 3.0196$ for $\epsilon = 10^{-3}$ using large-$t$ expansion of $\mean{R(t)}$. Corresponding relative depth of correlation hole is $\kappa \equiv 1 - \frac{ \mean{R(\tTh)} }{\overline{R}} = e^{-\frac{3}{2}} \approx 0.22313$.
\begin{table}[t]
	\begin{tabular}{|c|c|c|c|c|c|c|c|c|}
		\hline
		& $\mean{R(t\ll 1)}$           & $\Gamma$              & $\mean{R(t\gg 1)}$                                      & $\tTh$                                             & $R(\tTh)$      & $\tR$                                    & $\overline{R}$  &  $\kappa$ \\ \hline
		$N = 2$   & $1-\frac{t^2}{2}$ & $\sqrt{\frac{3}{2}}$ & $\overline{R} - \frac{1-\overline{R}}{2t^2}$ & 1.50198                                            & 0.678813       & $\sqrt{\frac{1}{6\epsilon}}$            & $\frac{3}{4}$  &  0.0949  \\ \hline
		$N \gg 1$ & $1 - Nt^2$         & $\sqrt{N}$            & $\overline{R} - \frac{1-\overline{R}}{3t^2}$ & $\del{\frac{3}{\pi}}^{\frac{1}{4}}\approx 0.9885$ & $\frac{2}{N}$ & $\frac{1}{3}\sqrt{\frac{N}{\epsilon}}$ & $\frac{3}{N}$  &  $\frac{1}{3}$ \\ \hline
	\end{tabular}
	\caption{Ensemble averaged fidelity in GOE: comparison between $N = 2$ and $N\gg 1$. $\Gamma$ is the width of the LDOS and $\kappa \equiv 1 - \frac{ \mean{R(\tTh)} }{\overline{R}}$ is the relative depth of correlation hole. $\tR$, the relaxation time is defined as the point where $\mean{R(t)}$ approaches the $\epsilon$-neighborhood of the equilibrium value, $\overline{R}$, where $\epsilon\ll 1$ is the tolerance value. For example, $\epsilon = 10^{-3}\Rightarrow \tR\approx 13$ for $N = 2$.}
	\label{tab_Rt_GOE_comparison}
\end{table}

\section{Change of variables for $2\times 2$ S-RPE}\label{sec_A_symp}
Normalized eigenvectors of a symplectic matrix have the form
\begin{equation}
	\label{eq_H_symp_vec}
	\ket{\Psi} = \frac{1}{2\sqrt{2\csc^2\zeta - 1}}\left(\begin{array}{c}
		-\cot\frac{\zeta}{2} \del{\sin\alpha\cos\theta + i\cos\alpha\sin\phi}\\
		-\cot\frac{\zeta}{2} \del{-\cos\alpha\cos\phi + i\sin\alpha\sin\theta}\\
		\tan\frac{\zeta}{2} \del{\sin\alpha\cos\theta + i\cos\alpha\sin\phi}\\
		\tan\frac{\zeta}{2} \del{-\cos\alpha\cos\phi + i\sin\alpha\sin\theta}
	\end{array}\right)
\end{equation}
\begin{table}[b]
	\raggedleft
	\begin{tabular}{|c|c|c|}
		\hline
		u	& v		& x \\
		\hline
		$\frac{(E_1+E_2) + (E_1-E_2)\cos\zeta}{2}$ & $\frac{(E_1+E_2) - (E_1-E_2)\cos\zeta}{2}$ & $\frac{E_1-E_2}{2}\sin\zeta\cos\alpha\cos\phi$ \\
		\hline
		y		& z		& w\\
		\hline
		$\frac{E_1-E_2}{2}\sin\zeta\cos\alpha\sin\phi$ & $\frac{E_1-E_2}{2}\sin\zeta\sin\alpha\cos\theta$ & $\frac{E_1-E_2}{2}\sin\zeta\sin\alpha\sin\theta$ \\
		\hline
\end{tabular}
\caption{Change of variables for $2\times 2$ symplectic matrix.}
\label{tab_change}
\end{table}
Then we can use the change of variables in table~\ref{tab_change} with Jacobian of transformation, $|J| = \frac{1}{32}(E_1-E_2)^4\abs{\sin^3\zeta\sin 2\alpha}$. Then for $2\times 2$ S-RPE, the density in matrix space can be written as
\begin{equation}
	\eqalign{
		&\prob{H} = \frac{1}{8\pi^3\sigma^4}\exp\del{-\frac{u^2+v^2}{2} - \frac{x^2+y^2+z^2+w^2}{2\sigma^2}}\\
	\Rightarrow &\prob{\vec{E}, \zeta, \alpha, \theta, \phi} = \frac{1}{2^8\pi^3\sigma^4} \abs{\sin^3\zeta \sin 2\alpha} \del{E_1-E_2}^4 e^{ - \frac{E_1^2 + E_2^2}{2} + \frac{(E_1-E_2)^2}{4}\sin^2\zeta\del{1 - \frac{1}{2\sigma^2}} }\\
	}
\end{equation}
where $\sigma^2 = \frac{1}{2^{\gamma+1}}$. Note that $\zeta$ here plays the role of $2\theta$ in case of O and U-RPE.
\pagebreak\newpage

\bibliographystyle{unsrt}
\end{document}